\documentclass[prl,twocolumn,floatfix]{revtex4}

\usepackage{amsmath}  
\usepackage{amsfonts}  
\usepackage{graphicx}   

\newcommand{\be}{\begin{eqnarray}}
\newcommand{\ee}{\end{eqnarray}}

\newcommand{\vk}{{\bf k}}
\newcommand{\vp}{{\bf p}}

\newcommand{\kB}{k_{\rm B}}

\newcommand{\dby}[2]{ \frac{{\rm d} #1}{{\rm d} #2}}

\newcommand{\myfig}[2]
{\centerline{\resizebox{!}{#1\textwidth}{\includegraphics{#2}}}}

\begin{document}

\title{Unruh effect and macroscopic quantum interference}

\author{Andrew M. Steane}
\affiliation{Department of Atomic and Laser Physics, Clarendon Laboratory, Parks Road, Oxford OX1 3PU, England.}

\date{\today}

\begin{abstract}
We investigate the influence of Unruh radiation on matter-wave interferometry experiments using
neutral objects modeled as dielectric spheres. The Unruh effect leads to a loss of coherence through
momentum diffusion. This is a fundamental source of decoherence that affects all objects having
electromagnetic interactions. However, the effect is not large enough to prevent the observation of
interference for objects of any size, even when the path separation
is larger than the size of the object. When the acceleration in the interferometer
arms is large, inertial tidal forces will disrupt the material integrity of the interfering objects
before the Unruh decoherence of the centre of mass motion is sufficient to prevent observable interference. 
\end{abstract}




\maketitle

There is wide interest in discovering whether or not quantum interference can be observed
for large objects. There are many ways in which this issue can be stated precisely.
One of the natural ways is simply to take a lump of ordinary matter
of larger and larger size, and discover whether or not interference fringes can be observed when
the lump is made to pass through two slits, or, more generally, pass through an
interferometer, the arms of which are separated by a distance larger than the diameter of
the lump. 
As the size of the lump grows, so does its mass, and therefore its de Broglie
wavelength at any given velocity falls. Consequently the interferometer gets more and
more sensitive to small effects which may randomize the interference phase---the problem
of decoherence. Therefore one expects to encounter limits which make
it not feasible to observe such interference above some size of the lump. The question
then arises, whether these limits are purely technological and could in principle be overcome, 
or whether the natural world itself poses intrinsic limits to quantum coherence. One may, for example,
conjecture that the natural free motion of all physical entities has a stochastic component,
or some other non-linear decohering process, not described by the
existing formulation of quantum mechanics (quantum field theory). Or, one may argue
that the existing standard model of physics already places fundamental limits to
quantum coherence. 

In this paper we address a mechanism which restricts the visibility of interference in matter-wave
interferometers, {\em within} the existing standard formulation of physics. That `standard formulation' we
take to be quantum field theory 
on a background of classical spacetime described by general relativity. The mechanism
we shall consider is Unruh radiation \cite{76Unruh,73Fulling,08Crispino,06DeBievre}. 
This may be said to be `fundamental', i.e. implicit in the laws of motion, and unavoidable. A related
phenomenon is the dynamical Casimir effect, in which the vacuum radiation pressure dissipates
the kinetic energy of a moving mirror \cite{00Dalvit}; we comment on that connection at the end
of the paper.

\begin{figure}
\myfig{0.18}{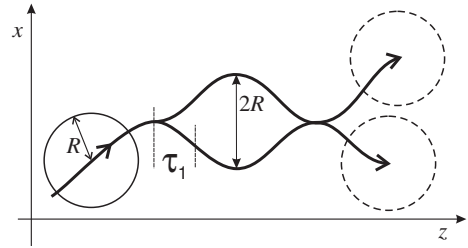}
\caption{A sphere of radius $R$ passes through a two-path interferometer whose
path separation is $2R$. The figure shows the paths in such an interferometer when the acceleration
is spread along the whole trajectory, such that the motion consists of concatenated periods
of constant proper acceleration. $\tau_1$ is the proper time between the beam splitting
and the event at which the proper acceleration first changes sign.}
\label{f.int}
\end{figure}

We consider a generic interferometer with two interfering `arms'.
We study interference of the de Broglie waves associated with a 
sphere of proper radius $R$, and consider the case where the separation of the
arms is $2R$ (see figure \ref{f.int}a). This captures the idea of a macroscopic
separation of paths, such that the interfering object might be said to have been
`separated from itself' in the middle of the interferometer.

The argument of the present
paper may be summarized as follows. Any matter-wave interferometer in which
the arms enclose a non-zero area of spacetime involves acceleration.
Owing to the Unruh effect, the accelerating object experiences a fluctuating force which
leads to momentum diffusion. This in turn leads to fluctuation of the interference phase of
the interferometer. We estimate the size of this effect for a wide class of physical objects.
We find that the blurring effect is there, but it does not does
not limit the size $R$ of object for which interference can be observed. Also, we find the
observation of interference does not place any new constraint on the acceleration of the
object, over and above
the one already imposed by the requirement that it is not torn apart by tidal inertial forces.

We now turn to this argument.

The Fulling-Davies-Unruh effect can be presented in more than one
way \cite{08Crispino,06DeBievre,84Unruh,05Fulling}. At the heart of
it is Unruh's observation that, in the coordinate system of the Rindler frame 
(constantly accelerating frame in flat spacetime), the vacuum state of quantum field theory
takes the form of a thermal state with temperature
\be
\kB T = \frac{\hbar a}{2\pi c} .
\ee
where $a$ is the proper acceleration.
This statement is made in a more thoroughly well-defined way in the literature, and
we shall elaborate on it shortly. There is not yet a complete consensus on the 
precise meaning and the physical implications \cite{04Fulling,05Fulling}, but it will be sufficient to our purpose
to take the following broadly standard point of view. We consider an {\em inertial} reference
frame, with the electromagnetic field initially in its vacuum state.
In this frame we suppose there exists a system A which provides a force on system B,
causing B to accelerate. B may be any system having electromagnetic interactions,
such as a charged body, a detector, a dipole, a lump of fused silica. In this situation,
we claim, the net force on B will fluctuate. 
If B has internal structure, then
it will undergo internal excitations, and subsequently emit photons, in a stochastic
way. The energy is provided by system A;
the stochastic nature of the process is owing to the fluctuating vacuum and therefore
is unavoidable. If the proper acceleration is constant on average 
and goes on for long enough, then
the fluctuation is the same as if B were bathed in thermal\footnote{The radiation is
`thermal' in the sense of Eqn (\ref{nbar}).} radiation 
at the Unruh temperature in its instantaneous rest frame. 
If B has no internal structure but is an accelerating
charged particle, the QED treatment of radiation reaction equally 
leads to a fluctuating force  \cite{91Ford,05Fulling}.

The importance of Unruh's calculation is that it suggests
the effect under consideration is owing to basic kinematics of the electromagnetic
field, and therefore is universal. The response of B is 
very like the one it would have if it were at rest and bathed in thermal radiation
at the Unruh temperature, and that temperature depends only on the acceleration, not on other
details of either A or B. We say `very like' rather than `identical to' because it is
not necessary to our argument that the two cases be identical, only that they be like.
Following Boyer \cite{80Boyer}, we make the following claim. A physical entity
(often called `observer') accelerating through the vacuum will undergo internal
excitations similar to those it would undergo if it were moving inertially and
subject to radiation such that the density matrix of the electromagnetic field is diagonal in
the Fock basis, with a mean excitation per mode
\be
\bar{n}(\omega) = \frac{1}{e^{2\pi c\omega/a}-1}
\left( 1 + 2 \left( \frac{a}{\omega c}\right)^2 \right).  \label{nbar}
\ee
This spectrum is not quite the ordinary thermal (Planck) form but is closely
related to it, and we shall refer to this radiation as `thermal' in the following.
We say the excitations of the accelerating entity are similar, not
identical, to those of the corresponding inertial entity because it is not possible
to make a more precise statement unless one investigates how the internal dynamics
of the entity are affected by acceleration. By arranging the forces so as to
accelerate B while keeping its internal stress to a minimum, one may arrange
that no large discrepancy will arise by this route, as long as 
the B is small enough for tidal effects to be negligible.

A further source of imprecision is the fact that the motion under consideration
will only involve acceleration for finite periods of time, so the Unruh result does
not apply exactly. Acceleration for finite periods is discussed in \cite{03Martinetti,08Crispino,08Barton}. The
approximation that the Unruh temperature applies to the majority of the elapsed
proper time is good when the product of proper acceleration and
proper time is of order $c$.

Now consider a generic matter-wave interferometer whose arms enclose a
non-zero area in spacetime. As we already commented, such an interferometer involves acceleration
and therefore the Unruh effect will come into play. The only way to avoid
this is if both arms of the interferometer are geodesic (that is, they both represent free
fall motion). This is in principle possible for a large part of the motion,
for example if the arms pass either side of
a massive gravitating object, but it is not clear whether or not 
the action of the beam splitters
must involve non-gravitational forces. In any case,
we will restrict the treatment to the case of flat spacetime in the following.

It will emerge that the decoherence scales as
a high power of $a$, and therefore to minimize the decoherence, a long period
of low $a$ is better than inertial motion combined with a short period of large $a$.
Therefore we will study a trajectory made of three periods of constant proper
acceleration, as shown in figure \ref{f.int} \footnote{Different parts of the sphere
here experience different amounts of proper time, and this will lead to further decoherence
unless the sphere is at zero temperature. To avoid this the sphere can be cooled, or
the interferometer can be extended to a more symmetric configuration involving
two areas traversed in opposite directions (sometimes called `zero area').}.
We model the interfering object as a dielectric sphere of relative permittivity $\epsilon$ at low
frequencies. Such a model applies to a wide class of objects as long as the dominant
wavevectors of the electromagnetic radiation under consideration satisfy $k R \ll 1$.
For the motion under consideration, the spectrum of the radiation is thermal
with typical wavevector $k \simeq k_{\rm B}T/\hbar c = a/2\pi c^2$ so the
model is valid when $aR/c^2 \lesssim 2\pi$; we will show at the end
that this condition holds.

According to our interpretation of the Unruh effect, the excitation of the sphere causes 
it to behave, in the instantaneous rest frame, 
as it would if it were scattering thermal radiation. When an incident photon
of wavevector $\vk$ is scattered, the momentum of the sphere changes by $\delta \vp = \hbar(\vk-\vk')$.
For a heavy sphere ($mc \gg \hbar k$), $k'=k$ and so  $\delta p^2 = \hbar^2 k^2 2(1-\cos \theta)$.
The photons arrive from random directions\footnote{We do not need to assume isotropic radiation here, only that the random component in the directions is well modelled by a random walk with the assumed step size.}
and therefore the momentum undergoes a random walk,
such that after proper time $\tau$ the momentum variance is $\Delta p_0^2 = \hbar^2 k^2 \Gamma_k \tau$
where \cite{85Joos}
\be
\Gamma_k = \int \dby{\sigma}{\Omega} {\rm d}\Omega \frac{\bar{n}c}{V} 2(1-\cos \theta)
\,= \frac{16\pi}{3} \frac{\bar{n}c}{V} k^4 R^6 \left(\! \frac{\epsilon-1}{\epsilon+2}\right)^{\!\!2} \label{Gammak}
\ee
Here we have used the classical cross-section for scattering by a dieletric sphere in the limit
$k R \ll 1$ \cite{98Jackson}. $V$ is the volume of space containing the electromagnetic field; it will go to
infinity at the end of the calculation when we integrate over $k$.
$\bar{n}$ is given by Eqn (\ref{nbar}) with $\omega=ck$.

The phase of the de Broglie waves is Lorentz-invariant and is most conveniently calculated in
an inertial frame $F$ that moves in the $z$ direction relative to the beam-splitters shown in figure
\ref{f.int}, such that the sphere moves along the $x$ axis of $F$.
In $F$, the momentum of the sphere
is related to that in the instantaneous rest frame by a Lorentz transformation. Consequently it is
distributed with a standard deviation given by $\Delta p= \gamma \Delta p_0 = \gamma \hbar k (\Gamma_k\tau)^{1/2}$ where $\gamma$
is the Lorentz factor.

In the absence of momentum diffusion, the phase accumulated along one
interferometer arm is given by a path integral along the classical trajectory $x(\tau)$. 
In the presence of momentum diffusion, the phase gradually acquires a spread given by
\cite{83Caldeira,90Stern}
\be
\Delta\phi_k &=& \int \frac{\Delta p \,{\rm d}x}{\hbar} 
\;=\; 
\int \frac{\gamma \Delta p_0}{\hbar} \left| \dby{x}{\tau} \right| {\rm d}\tau  \nonumber \\
& = &  k \int \gamma^2 |v| \sqrt{\Gamma_k \tau} \,{\rm d}\tau    \label{dphik}
\ee

\begin{figure}
\myfig{0.2}{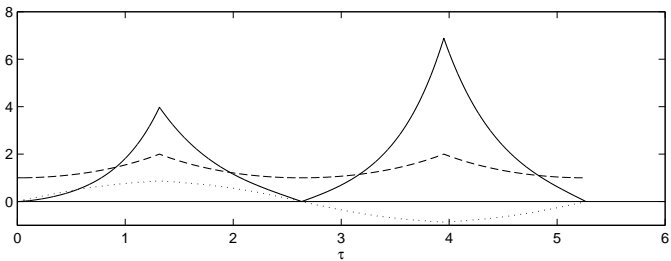}
\caption{The integrand in eqn (\protect\ref{dphik}), for the case $a=1,\;R=2,\;c=1$.
Dots: $v$; dashes: $\gamma$; full curve: $\gamma^2|v| \tau^{1/2}$. }
\label{f.integ}
\end{figure}

Let $\tau_1$ be the duration of the first period of constant proper acceleration. The
worldline is given by $x = x_0 + (c^2/a) \cosh(a\tau/c)$ with $x(\tau_1)-x(0)=R/2$
hence 
\be
\tau_1 = \frac{c}{a} \cosh^{-1}\left(1 + \frac{aR}{2c^2}\right) 
\;\;\;\;\;\;\;\;\; \left(\simeq \sqrt{R/a} \right)  \label{tau1}
\ee
and $\gamma = \cosh(a(\tau-2j\tau_1)/c),\; v=c\tanh(a(\tau-2j\tau_1)/c)$ where $j=\{0,1,2\}$
for the three parts of the worldline between the beam splitters, of proper duration
$\tau_1,\,2\tau_1,\,\tau_1$ respectively.
The integrand in Eqn (\ref{dphik}) is shown in figure~\ref{f.integ}. The integral is
$k c\sqrt{\Gamma_k c^3/a^3}$ multiplied by a function of $(a\tau_1/c)$ which
we obtained by numerical integration. Since the Unruh effect is only expected when
the acceleration goes on for long enough, such that $a \tau_1$ is of order $c$, 
we are only interested in studying the integral for values of $a\tau_1/c$
in the range $0.5 < a\tau_1/c < \cosh^{-1}(3/2)$. In this range the result can
be approximated, to $0.15$\% accuracy, by
\be
\Delta \phi_k^2 \simeq 7.325 \, \Gamma_k \tau_1 \left(\frac{k c^2}{a}\right)^{\!2} 
\sinh^4\left(\frac{a \tau_1}{c}\right).
\ee
Here we have also introduced the approximation that $\Gamma_k$ is independent of
$\tau$, which means we ignore the complications in the Unruh effect 
associated with a finite period of acceleration. The formula for the cross-section in
Eqn (\ref{Gammak}) is also approximate since the sphere is accelerating. The assumption
of rigid motion (constant proper dimensions) makes this a reasonable first approximation, but
a more thorough analysis would be needed to check the degree of approximation involved.
Using  (\ref{tau1}), we have
\be
\Delta \phi_k^2 \simeq 7.3 \, \Gamma_k \tau_1 (k R)^2 \left(1 + \frac{aR}{4 c^2}\right)^2.
\label{dphik2}
\ee

So far we have obtained the variance of the phase owing to fluctuations at one frequency.
Using (\ref{nbar}), (\ref{Gammak}) and (\ref{dphik2}), 
the variance owing to fluctuations at all frequencies is
\be
\!\Delta \phi^2 &=& \int \Delta \phi_k^2 \frac{V {\rm d}^3 k}{(2\pi)^3}   \nonumber \\
&\simeq& 0.04 \left(\! \frac{\epsilon-1}{\epsilon+2}\right)^{\!\!2} 
\left(1 + \frac{aR}{4c^2}\right)^2 \frac{\tau_1 c}{R}  \left( \frac{a R}{c^2} \right)^9\!\!.
\label{dphi2}
\ee
Here, we took $\epsilon$ as independent of frequency, which is valid for an ordinary material since
we assumed $k R \ll 1$ and we are interested in values of $R$ of the order microns or larger. Therefore
$\epsilon$ is the low-frequency (d.c.) relative permittivity.

Finally, we need to consider the combination of both arms of the interferometer. If the interferometer
were really bathed in thermal radiation, then the low-frequency contribution to the fluctuations in the two arms 
would be correlated. However, here there is no
incoming radiation (in the Minkowski frame). Rather, the scattering calculation is a mathematical method that is
being used to estimate fluctuations of the forces that are causing acceleration in the two arms. 
Since these forces are acting in spacelike separated regions it is not self-evident that their fluctuations would
be correlated, but in view of the fact that the effect involves the quantum vacuum one cannot rule out
correlation. This is an open question. Here we shall treat the case of no correlation; in this case
when both arms of the interferometer are included, the interference phase uncertainty
is $\Delta \phi_{\rm tot} = \sqrt{2} \Delta \phi$. Correlations would
be expected to result in less decoherence, so we thus obtain an upper bound on the decoherence. We note
that our final result has some similarity with a prediction for the dynamical Casimir effect (DCE) considered below,
which suggests that the Unruh effect does lead to decoherence.

Now, we have assumed all along that the sphere is substantially unaffected by its acceleration.
For example, one may imagine that the forces on it are arranged such that it retains
fixed proper dimensions. However, if $a$ is the proper acceleration of the center of the sphere,
then when $R = c^2/a$ the surface of the sphere extends to the horizon in Rindler space. In
other words, this is the condition where the sphere can no longer move rigidly and tidal forces
are large: not merely non-negligible but insurmountable. At this limiting value of $R$ we find
\be
\Delta \phi_{\rm tot} \simeq 0.35 \left( \frac{\epsilon-1}{\epsilon+2} \right).
\ee 
Hence we find that $|\Delta \phi_{\rm tot}| \ll \pi$ for $\epsilon > 0$. This means the interference
fringes will be easily visibile for
any positive $\epsilon$. We therefore conclude that the decoherence owing
to the Unruh effect is small: it is not sufficient to prevent interference even when the acceleration
is so extreme that the sphere begins to break up.
Schr\"odinger's cat would be killed by the
inertial forces before its acceleration was sufficient for this form of decoherence to be substantial.
Also, owing to the high power of $(aR/c^2)$ in Eqn. (\ref{dphi2}), the decoherence is very
small for any value of $R$ less than $c^2/2a$. But this condition is, within a
numerical factor of order 1, the same as the one required by purely classical relativistic
considerations if the interfering entity is not to
experience extreme tidal forces. 

We now compare the above with the conclusions for DCE as described in \cite{00Dalvit}.
In the case of an oscillating mirror, Dalvit and Maia Neto find that the effect of DCE is to single out a
pointer basis consisting of
coherent states of mirror motion (c.f. \cite{04Eisert}), and the off-diagonal elements of the density 
matrix for a superposition of coherent states decay at a rate $\Gamma$.
In the perfectly reflecting limit, this decoherence rate is given by
$\Gamma =  |\alpha|^2  \hbar \omega_0^2 / 3 \pi M c^2$
where $\alpha$ is a coherent state parameter, $M$ is the mass of the mirror and $\omega_0$ is
the oscillation frequency. The calculation is carried out in the low velocity limit, $v \ll c$.
To compare this with our results, consider the case where the amplitude of the oscillation 
associated with the coherent state is $R$. 
Then $|\alpha|^2 = R^2 M \omega_0/2\hbar$ so $\Gamma = R^2 \omega_0^3/6\pi$. The situation
comparable to our treatment above is when the mirror can oscillate for a half-period without losing
coherence. Suppose the mirror can oscillate for $N$ half-periods before substantial decoherence occurs.
The condition for this is  $\Gamma \lesssim \omega_0 / N \pi$, which gives $R^2 \omega_0^2 \lesssim 6 c^2 / N$.
Using that the acceleration is of order $a \simeq \omega_0^2 R$, this can be expressed
$R a /c^2 \lesssim 1 / N$. For $N=1$ this agrees with our conclusions above, although 
the two calculations are very
different, one involving a general estimate of the fluctuations and treating $v \sim c$, the other
employing field theory for a more specific system in the limit $v \ll c$. 

With the benefit of hindsight, one might claim that the limiting condition on $(aR)$ could be obtained by dimensional analysis, but it was
not self-evident at the outset that Planck's constant would not appear in the result, and indeed this simple
condition was not remarked in \cite{00Dalvit}. Also, it emerged in the latter case in a calculation in
the limit $v \ll c$ with no role for Special Relativity as such.
Of course for $N=1$ the DCE calculation is not valid near the upper bound, since the condition then gives
$v \simeq c$, and our Unruh effect calculation is not valid for $a \tau_1 \ll c$. The DCE calculation 
does not exhibit the strong scaling with $(aR/c^2)$ that we observe in eqn (\ref{dphi2}).
This implies that the DCE is only loosely related to the Unruh effect, 
or else that a qualitative change in the latter occurs when one passes from $a \tau_1 \sim c$
to $a \tau_1 \ll c$.


If ones interest is in observing the influence of the Unruh effect, one could use particles with
a large charge to mass ratio, such as single electrons,
and then a much larger effect will be found \cite{99Chen,01Breuer}. 
An enhancement might also be available by the use of plasmonic material or an
exotic material with a relative permittivity approaching $-2$ over a significant frequency range. 
In this case one must allow for the frequency-dependence when performing the integral in eqn (\ref{dphi2}).
We have instead studied the question, whether or not the effect is large enough
to prevent observable interference of ordinary, large, neutral objects when the interferometer is designed
to avoid acceleration as much as possible. Our conclusion is that the Unruh effect
places no limit on the size of a cold lump of ordinary matter that might in principle be made to exhibit
interference after passing either size of an obstacle.

I thank Jasper van Wezel for helpful comments and discussion.

\bibliography{selfforcerefs}

\end{document}